# Does the Hubble Redshift Flip Photons and Gravitons?

Matthew R. Edwards

*Gerstein Science Information Centre, University of Toronto, Toronto, Ontario, Canada, M5S 3K3*

*e-mail:* matt.edwards@utoronto.ca

18 pages, 2 tables, 1 figure

**Abstract**. Due to the Hubble redshift, photon energy, chiefly in the form of CMBR photons, is currently disappearing from the universe at the rate of nearly $10^{55}$ erg s$^{-1}$. An ongoing problem in cosmology concerns the fate of this energy. In one interpretation it is irretrievably lost, *i.e.*, energy is not conserved on the cosmic scale. Here we consider a different possibility which retains universal energy conservation. If gravitational energy is redshifted in the same manner as photons, then it can be shown that the cosmic redshift removes gravitational energy from space at about the same rate as photon energy. Treating gravitational potential energy conventionally as 'negative', it is proposed that the Hubble shift flips positive energy (photons) to negative energy (gravitons) and *vice versa*. The lost photon energy would thus be directed *towards* gravitation, making gravitational energy wells more negative. Conversely, within astrophysical bodies of sufficient size, the flipping of gravitons to photons would give rise to a 'Hubble luminosity' of magnitude $-UH$, where $U$ is the internal gravitational potential energy of the object and $H$ the Hubble constant. Evidence of such an energy release is presented in bodies ranging from planets, white dwarfs and neutron stars to supermassive black holes and the visible universe.

## 1. Introduction

As the universe expands in the standard cosmology, the number density of photons within each expanding volume of space remains constant. The wavelength of each photon at the same time is increased and so there is a net loss of photon energy in the universe. The largest pool of photon energy is the cosmic microwave background radiation (CMBR) and the loss of energy from this pool due to the redshift is readily estimated at about $10^{55}$ erg s$^{-1}$. This is roughly equal in magnitude to the total luminosity of all the stars in the visible universe. Not surprisingly, an important question in cosmology concerns the fate of this energy. Harrison (1995) observed that the energy does not apparently go into perturbations of the spacetime metric, since those disturbances, as they propagate, would also lose energy because of the cosmic redshift. The question would then become where did the gravitational energy go? With respect to the lost photon energy Harrison thus concluded: "Does the energy totally vanish, or does it reappear, perhaps in some global dynamic form? The tentative answer based on standard relativistic equations is that the vanished energy does not reappear in any other form, and therefore it seems that on the cosmic scale energy is not conserved."

Returning to Harrison's abandoned suggestion that the lost photon energy could reappear within the spacetime metric, let us consider the role of gravitational potential energy. Gravitational potential energy is an enigmatic concept in both newtonian and relativistic physics. In the former it cannot be localized to any specific part of the system, while in the latter the whole concept of potential energy is lacking definition. Yet, if gravitational potential energy takes the form of discrete waves within the metric, then it can easily be shown that the universe's stock of it would suffer an almost equal energy rate of loss ~ $10^{55}$ erg s$^{-1}$. There would then be not one but two pools of energy steadily dissipating at the same enormous rate. If these processes were unconnected, it would be a remarkable coincidence.

Treating gravitational potential energy conventionally as negative energy, it has been proposed by the author that the Hubble shift 'flips' photons (positive energy) to gravitons (negative energy) and *vice versa*. In the former instance, the lost photon energy would be directed towards gravitation, making gravitational energy wells more negative, while concurrent graviton flipping would return photon energy to the universe. Balance between the cosmic pools of gravitational and electromagnetic energy would thus be maintained and the total energy of the universe held constant.

In this paper we will not attempt to develop the theoretical foundations for this idea, which clearly would assume a complex and unconventional form. Rather, we will be concerned with the more mundane question: does it work or not? To test it, it would be necessary to show that new gravitons are being formed out of photons and that new photons are also being generated from gravitons. The detection of new gravitons is problematic on several levels, regardless of their mode of synthesis. However, the reverse process of photon production from gravitons is potentially quite observable. This is because photons would be generated not only between discrete gravitating masses but within masses as well. All bodies, from a grain of sand to the universe as a whole, would be subject to a heating effect due to conversion of internal gravitational energy to photons. This process would constitute a new luminosity input in all bodies, which we now term the Hubble luminosity, $L_H$. $L_H$ is defined as the luminosity that a distant observer would measure in a body if all the energy derived from conversion of its gravitational potential energy to photons via the model process were radiated away as heat. Its magnitude is given by

$$L_H = -UH_0, \tag{1}$$

where $U$ is the internal gravitational potential energy and $H_0$ is Hubble's constant.

The hypothesis of photon/graviton flipping mediated by the cosmic redshift was earlier discussed with reference to planetary heat emissions and white dwarf luminosities (Edwards, 2006, 2008). The excess heat emissions of Earth, Saturn, Neptune and Jupiter amount to 5-10 per cent of their respective Hubble luminosities. For these bodies, the remaining 90-95 per cent of the released energy could potentially be involved in diverse processes such as phase changes, tectonics or planetary expansion. The luminosities of a sample of white dwarfs for which mass and radius are independently estimated, *i.e.,* without using the white dwarf mass-radius relationship, were also shown to be within an order of magnitude of their Hubble luminosities. In this paper we extend the analysis to include neutron stars, supermassive black holes and to the visible universe.

It is important to note that the model differs fundamentally from earlier suggestions by Dirac (1937) and others that the gravitational constant $G$ decays at a fractional rate proportional to $H_0$, for which observational evidence has been mostly negative (Uzan 2003). Due to continuous regeneration of gravitons from photons in the model, gravitational forces between bodies, as well as $G$, do not diminish over time.

## 2. Evidence from White Dwarfs, Neutron Stars and Black Holes

Since the internal gravitational potential energy of a star or planet is proportional to $M^2$ and inversely to $R$, the best opportunities to test the model are with very large and compact bodies. For main sequence stars, like the sun, the energy released in graviton decay would be small compared to stellar fusion. In the case of the sun, the quantity $-UH$ amounts to $5 \times 10^{30}$ erg s$^{-1}$, almost three orders of magnitude lower than its actual luminosity. However, for stars that are either too small to have initiated sustained fusion (*e.g.*, small brown dwarfs) or now conduct little fusion and moreover are compact (*e.g.*, white dwarfs, neutron stars), the Hubble luminosities are much more potentially measurable. While insufficiently precise data is available yet on small brown dwarfs, the situation is better for white dwarfs and neutron stars. We first briefly review the results of our white dwarf study, before considering neutron stars and supermassive black holes.

### 2.1 White Dwarfs

White dwarfs have radii similar to Earth's and masses less than 1.4 $M_\odot$. Due to their very high densities (~ $10^6$ g cm$^{-3}$), the electrons in white dwarfs are degenerate and this provides the theoretical basis for the WD 'mass-radius relationship'. White dwarfs (WDs) begin their lives with very high luminosities and are considered to gradually cool off and fade away over a period of 5-10 billion years. The main contributor to their luminosity is considered to be the loss of thermal energy acquired during gravitational collapse from the precursor star. For white dwarfs with $T_{eff}$ > 12,000 K, radiative cooling through the surface non-degenerate layer is the dominant mode of cooling. Below this temperature, a convection layer forms which increases in thickness as the white dwarf cools. The very low temperatures of white dwarfs with $T_{eff}$ ≤ 5,000 K are considered to reflect the ages of the white dwarfs and thus, in the conventional interpretation, potentially provide constraints on the ages of the galactic disk and the universe.

For white dwarfs with masses much less than the Chandrasekhar limit of ~ 1.4 $M_\odot$, the quantity $U$ is conventionally given as

$$U \equiv -\frac{GM^2}{R}. \tag{2}$$

The Hubble luminosity $L_H$ is then

$$L_H = \frac{GM^2}{R} H. \tag{3}$$

In their efforts to study the mass-radius relationship itself, Provencal *et al.* (1998) and later Mathews *et al.* (2006) analyzed a group of 21 white dwarfs for which mass and radius had been independently determined, *i.e.*, without using the WD mass-radius relationship. This same group was used in our study to test the model (Edwards, 2008). The total luminosity, $L$, for each star is obtained from $4\pi R^2 \sigma T_{eff}^4$, where $\sigma$ is the Stefan-Boltzmann constant (= $5.67 \times 10^{-5}$ erg cm$^{-2}$ (deg-K)$^{-4}$ s$^{-1}$). The results are shown in Table 1. For the 16 hotter DA white dwarfs in the sample ($T_{eff} \geq 12,000$ K), the predicted luminosities are well within an order of magnitude of the observed luminosity, with a mean ratio $L/L_H = 0.84$. For the five cooler white dwarfs – G226-29, L268-92, Procyon B, L481-60, and G156-64 – the observed luminosity is only 1-10% of $L_H$. However, later studies on these stars have sharply revised estimates of mass and/or radius (and through that $L$). When those estimates were used, four of the five stars were found to possess luminosities within an order of magnitude of $L_H$. Only Procyon B, a star already well-known for having strange properties, fell outside the model prediction.

## 2.2 Neutron Stars

There are many classes of neutron stars (magnetars, pulsars, *etc*.) and frequently they occur in binary systems. The binary stars often have a major contribution to their luminosity through accretion of infalling matter. For neutron stars it is thus necessary to focus on isolated neutron stars, for which this input is minimized. The cooling curves of isolated neutron stars share many features with those of white dwarfs. Only theoretical estimates exist for their radii and these are in the range of 9-16 km. The masses are more precisely constrained and are in the range of 1-2 $M_\odot$.

Data from a study of isolated neutron stars by Kaminker *et al*. (2006) was used to test the model. Those authors developed theoretical cooling curves for neutron stars of different masses and compared them to the observed luminosities of some isolated neutron stars. The bolometric luminosities were determined using

$$L_s^\infty = 4\pi\sigma R_\infty^2 (T_s^\infty)^4 , \qquad (4)$$

where $L_s^\infty$ is the surface thermal luminosity, $T_s^\infty$ the effective surface temperature and $R_\infty$ = $R/\sqrt{1 - 2GM/c^2 R}$ the apparent radius, all as seen by a distant observer. Independent estimates of $R_\infty$ and $M$ for the sample stars were not available. Kaminker *et al*. used $M$ = 1.4 $M_\odot$ in each case and allowed $R_\infty$ to vary between 11 and 16 km. The luminosities $L_s^\infty$ that they obtained were then found to be evenly distributed in the range of $10^{32}$-$10^{34}$ erg s$^{-1}$ (Table 2).

As in the white dwarf case, the Hubble luminosity is given by $L_H = GM^2H/R$, where $R$ is the real radius rather than the apparent radius, $R_\infty$. For stars with mass 1.4 $M_\odot$, $R_\infty$ is seen to be $\approx R + 3$ km. Using the approach of Kaminker *et al*., the values of $R$ for the sample would then fall in the range of 8-13 km. For a star with $M$ = 1.4 $M_\odot$ and $R$

= 8 km, we find that $L_H = 1.4 \times 10^{34}$ erg s$^{-1}$ and for a star with the same mass and $R = 13$ km, we have $L_H = 8.8 \times 10^{33}$ erg s$^{-1}$. The Hubble luminosities for neutron stars are thus seen to be narrowly peaked near $10^{34}$ erg s$^{-1}$. From this analysis we find that the ratio $L/L_H$ for neutron stars is in the range .01-1. The estimates of $T_s^\infty$ and $R_\infty$ used by Kaminker *et al.* to determine $L_s^\infty$ all involve numerous interpretations and assumptions, just as in the WD case. Nonetheless, despite these significant uncertainties, the observed luminosities once again appear to closely match the Hubble luminosities, in most cases being within an order of magnitude.

### 2.3 Supermassive Black Holes

Supermassive black holes (SMBHs) have masses in the range of $10^6$-$10^9$ $M_\odot$. They are now considered to reside at the centres of most if not all galaxies and are regarded as the power sources of active galactic nuclei (AGNs) and quasars. SMBHs are thought to have likely been present at the time when their respective galaxies were formed. The greatest part of the gravitational energy released during the formation of SMBHs should have been released billions of years ago. SMBHs thus serve as an important test of the model.

The Hubble luminosity of black holes is influenced heavily by relativistic effects. The radius is constrained near the Schwarzschild radius, which is defined as

$$R_S = \frac{2GM}{c^2}. \tag{5}$$

When inserted in equation 3, the model luminosity for a black hole then simplifies to

$$L_H = \frac{Mc^2 H}{2}. \tag{6}$$

It scales directly with the mass.

In the last decade surveys have been done on hundreds of AGNs which permit estimates of the mass and luminosity of their central SMBHs. Kollmeier *et al*. (2006) performed a study of 407 AGNs in the redshift range $z \sim 0.3\text{-}4$ and bolometric luminosity range $L_{bol} \sim 10^{45}\text{-}10^{47}$ erg s$^{-1}$. They found that the luminosities were sharply peaked at a value ¼ of the Eddington luminosity $L_{Edd}$, the value beyond which a black hole is unstable. $L_{Edd}$ is given conventionally as $L_{Edd} = 1.2 \times 10^{38} (M/M_\odot)$ erg s$^{-1}$. Comparison with equation 6 indicates that in black holes the ratio of the Hubble luminosity to $L_{Edd}$ is a fixed quantity, $L_H/L_{Edd} \approx 0.02$. Since $L_{bol}$ for the sample is peaked at $.25\, L_{Edd}$, it is seen that Hubble luminosities are nearly within an order of magnitude of the bolometric luminosities, with $L_H/L_{bol} \approx 0.1$.

To further test the model, we focused on the 19 nearest AGNs in their study in the range $z < 0.5$, for which the possible effects of redshift on measurements is presumably minimized. For these AGNs the masses were estimated to be evenly distributed in the range of $10^7\text{-}10^8\, M_\odot$ and the luminosities very sharply peaked at $L_{bol} \approx 10^{45}$ erg s$^{-1}$. The Hubble luminosity for a SMBH of $10^7\, M_\odot$ is found to be $L_H = 2 \times 10^{43}$ erg s$^{-1}$ and for one of $10^8\, M_\odot$ it is $L_H = 2 \times 10^{44}$ erg s$^{-1}$. For the nearest 19 AGNs, the Hubble luminosities are thus seen to evenly distributed in the range $.05 - 0.2\, L_{bol}$. Thus we find again $L_H/L_{bol} \approx 0.1$.

In the conventional interpretation, AGN luminosities are due to the gravitational energy released by infall of stars and other matter on the central SMBH. The absence of luminosities above $L_{Edd}$ is to be expected in any model, since the black holes would then be unstable. But that they should be so close to $L_{Edd}$, with almost none being less than $L_{bol} = .1\, L_{Edd}$, was not expected. Kollmeier *et al*. suggested that some form of black hole self-regulation may account for this. While quite conceivable, the present model predicts these luminosities.

### 3. Hubble Luminosity of the Universe

As inferred at the outset of this paper, radiation derived from the Hubble luminosity of the universe could replace the photon energy lost in the cosmic redshift, thus maintaining universal conservation of energy. We now consider this in greater detail. For a series of concentric shells of star-filled space of radius $r$ stretching outwards from a mass to the edge of the visible universe, the mass of each shell is proportional to $r^2$ while the quantity $U$ for each shell relative to the central mass is proportional to $1/r$. We thus find that for each shell $U \propto r$. As has long been recognized, the most remote shells of matter thus contribute by far the greatest portion towards the total gravitational potential energy of the central mass.

Let us assume that a mass can be influenced only by matter within the visible universe, which we will take as $R_U \approx c/H = 13.7$ billion light-years $= 1.3 \times 10^{28}$ cm. Since a sphere of radius $10^{28}$ cm has a volume one thousand times greater than the sphere of radius $10^{27}$ cm, we can regard all the universe's mass as effectively lying at the distance $R_U$. We find that the universal Hubble luminosity, designated $L_U$, is then

$$L_U = -UH = \frac{GM_U^2 H}{R_U} \ . \tag{7}$$

Expressing $M_U$ in terms of the universal density of matter via $V\rho = 4/3\pi R_U^3 \rho$, we have

$$L_U = \frac{G\rho^2 V_U^2 H}{R_U} , \tag{8}$$

and

$$L_U = \frac{16 G \rho^2 \pi^2 R_U^5 H}{9} \ . \tag{9}$$

For $R_U = 1.3 \times 10^{28}$ cm, $\rho = 10^{-31}$ gm cm$^{-3}$ (for baryonic matter only) and again $H = 2.2 \times 10^{-18}$ sec$^{-1}$, we have $L_U = 10^{55}$ erg sec$^{-1}$. Other hypothesized forms of mass (dark matter, dark energy, *etc.*) would increase this value, but may be superfluous if the observations supporting those forms are yet explicable through modifications to gravitational laws.

We next consider the photon energy lost due to the cosmic redshift. The largest known store of photon energy in the universe is the cosmic microwave background radiation (CMBR) with energy density $\rho_E \cong 4 \times 10^{-13}$ erg cm$^{-3}$. This radiation is being redshifted too and the universal rate of energy loss from the CMBR is thus $\rho_E V_U H = 4/3\pi(R_U)^3 \rho_E H$. Inserting the same values as above, this rate of photon energy loss is $8 \times 10^{54}$ erg sec$^{-1}$. Thus, if only baryonic matter is included in our estimate of $\rho$, the universal photon energy input through $L_U$ very closely matches the photon energy lost through the cosmic redshift.

If the replacement energy for the CMBR is taken as indeed a reflection of the Hubble luminosity of the visible universe, we would then see possible evidence of Hubble recycling in objects over a huge range of masses, from planets all the way to the universe as a whole. As shown in Fig. 1, the associated Hubble luminosities range over 36 orders of magnitude. The estimates of mass, radius and bolometric luminosity are subject to potentially large errors in the classes we have considered, in some cases greater than an order of magnitude. To truly test the validity of the hypothesis of graviton-photon flipping, it would be desirable to examine smaller objects, such as very small brown dwarfs or asteroids.

Due to the pressure of photons formed from graviton flipping, repulsive forces should arise between masses. For local systems, however, these can be shown to be negligible in comparison to the gravitational force. Let us suppose that in a system of two masses the product photons of graviton decay travel back along the axis connecting the masses and strike each mass. The repulsive force between the two masses is then

$$F = \frac{Gm_1 m_2 H}{Rc}. \qquad (10)$$

The factor $c$ appears in the denominator since a quantum of radiation with energy $E$ has momentum $E/c$. In the Earth-Moon system, for example, the repulsive force on the Moon results in an acceleration of only $\sim 10^{-18}$ cm s$^{-2}$, which is entirely negligible in comparison to the gravitational attraction. The two forces are seen to balance only when $r \approx R_U \approx c/H$.

## 4. Conclusions

In this paper we have considered whether the loss of photon energy through the cosmic redshift can be balanced by concurrent flipping of gravitons to photons and *vice versa*. Previously, this notion was examined with respect to planets and white dwarfs. For those classes of objects, the observed excess heat emissions/thermal luminosities were found to be generally within an order of magnitude of their respective Hubble luminosities. In the present paper evidence for this same pattern was also found in isolated neutron stars, supermassive black holes and in the universe as a whole. Preliminary evidence for the hypothesis of graviton flipping is thus seen to exist in the luminosities of masses scaling over 35 orders of magnitude. Though a formal treatment within GR is not yet available, we nonetheless conclude that graviton/photon flipping via the cosmic redshift appears to have the potential not only to address the problem of cosmic energy balance, but also to serve as an important new source of energy in astrophysical and geophysical processes generally. Future efforts to test the model more stringently should focus on smaller objects, such as asteroids and very small brown dwarfs, lacking significant radioactivity or fusion sources of energy.

| Object | $L$ ($\times 10^{31}$ erg s$^{-1}$) | $L_m$ ($\times 10^{31}$ erg s$^{-1}$) | $L/L_m$ |
|---|---|---|---|
| Sirius B | 9.0 | 39 | .23 |
| G226-29 | 0.77 | 18 | .043 |
| G93-48 | 7.7 | 13 | .59 |
| CD-38 10980 | 18 | 15 | 1.2 |
| L268-92 | 1.5 | 11 | .14 |
| Procyon B | 0.19 | 9.7 | .016 |
| Wolf 485 A | 3.0 | 7.7 | .39 |
| L711-10 | 9.4 | 7.3 | 1.3 |
| L481-60 | 0.81 | 7.7 | .11 |
| 40 Eri B | 5.0 | 6.1 | .82 |
| G154-B5B | 2.2 | 5.4 | .41 |
| Wolf 1346 | 9.9 | 4.8 | 2.1 |
| Feige 22 | 8.6 | 4.1 | 2.1 |
| GD 140 | 5.6 | 24 | .23 |
| G156-64 | 0.11 | 10 | .011 |
| EG 21 | 3.1 | 9.7 | .32 |
| EG 50 | 7.2 | 7.9 | .91 |
| G181-B5B | 1.4 | 7.5 | .19 |
| GD 279 | 1.9 | 4.9 | .38 |
| WD2007-303 | 3.0 | 5.0 | .60 |
| G238-44 | 8.2 | 4.9 | 1.7 |

**Table 1**. Model relationships. The second column gives the observed luminosity, the third column the model prediction and the last column the ratio of the observed luminosity to the model luminosity (from Edwards, 2008).

| Neutron Star | log $L_s^\infty$ (erg s$^{-1}$) |
|---|---|
| PSR J0205+6449 (in 3C 58) | <33.29 |
| PSR B0531+21 (Crab) | <34.45 |
| RX J0822−4300 | 33.9–34.2 |
| 1E 1207.4−5209 | 33.67–34.20 |
| RX J0007.0+7303 (in CTA 1) | <32.54 |
| PSR B0833−45 (Vela) | 32.19–32.67 |
| PSR B1706−4 | 31.66–32.94 |
| PSR J0538+2817 | 32.32–33.33 |
| PSR B0633+1748 (Geminga) | 31.34–32.37 |
| RX J1856.4−3754 | <32.5 |
| PSR B1055-52 | 32.05–33.08 |
| RX J0720.4−3125 | 31.37–32.40 |

**Table 2**. Surface thermal luminosity $L_s^\infty$ of some isolated neutron stars. From Kaminker *et al.*, 2006.

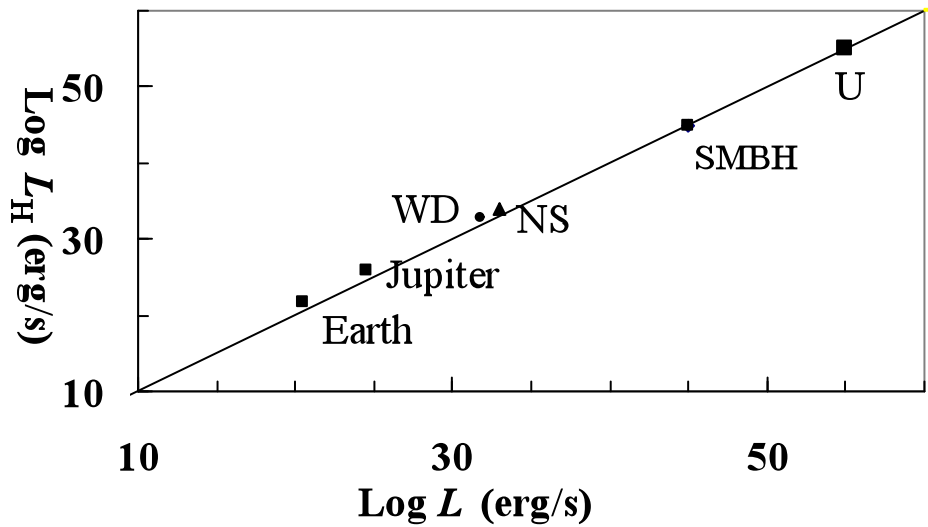

**Figure 1**. $L_H$ vs. $L$ at many scales. On the horizontal axis are plotted the bolometric luminosities of representative white dwarfs, neutron stars and supermassive black holes; the inferred luminosity of the universe from energy replacement; and the excess heat emissions of some planets. Their respective Hubble luminosities are on the vertical axis. The solid line is the 1:1 correspondence. The range of luminosities covers 35 orders of magnitude. Abbreviations: WD, white dwarf; NS, neutron star; SMBH, supermassive black hole; U, the universe.